\documentclass[seceq]{ptptex}

\title{
STOCHASTIC PROPERTIES OF THE FROBENIUS-PERRON OPERATOR%
}

\author{
Stephan I. \textsc{Tzenov}%
}

\inst{
CCLRC Daresbury Laboratory, Accelerator Science and Technology
Centre (ASTeC), Daresbury, Warrington, Cheshire, WA4 4AD UNITED
KINGDOM}

\abst{
In the present paper the Renormalization Group (RG) method is
adopted as a tool for a constructive analysis of the properties of
the Frobenius-Perron Operator. The renormalization group reduction
of a generic symplectic map in the case, where the unperturbed
rotation frequency of the map is far from structural resonances
driven by the kick perturbation has been performed in detail. It
is further shown that if the unperturbed rotation frequency is
close to a resonance, the reduced RG map of the Frobenius-Perron
operator (or phase-space density propagator) is equivalent to a
discrete Fokker-Planck equation for the renormalized distribution
function. The RG method has been also applied to study the
stochastic properties of the standard Chirikov-Taylor map. }

\begin{document}

\maketitle

\section{Introduction}
Recursive maps represent a useful and powerful tool to model and
to facilitate the understanding of the physical processes taking
place in complex nonlinear systems. In particular, they are widely
used to study the various transition scenarios from regular to
chaotic behaviour in nonlinear dynamical systems
\cite{Feigenbaum,Crutchfield,Manneville}, to simulate physical
systems exhibiting anomalous diffusion \cite{Geisel}, or to
analyze the underlying dynamics in time series with $1 / f$ noise
in their power spectrum \cite{Hirsch,Geisel1}. Iterative maps
provide a convenient and effective method to investigate
single-particle dynamics in accelerators and storage rings
\cite{Tzenov,Dragt}.

The extremely complicated behaviour of specific trajectories in
chaotic systems strongly suggests a probabilistic approach to the
dynamics. Instead of tracing an individual trajectory in phase
space, one employs a statistical mechanics approach by means of a
distribution function of an ensemble of trajectories. The
Frobenius-Perron operator of a phase-space density (distribution)
function, which sometimes is called the Transfer Operator of that
function or a phase-space density propagator, provides a tool for
studying the dynamics of the iteration of the distribution
function itself. The iterative map yields complete information of
how the value of an individual phase-space point jumps around
during successive iterations, so that one gets a good sense of the
point dynamics but no sense of how iteration acts on densities
with support on sets in phase space. The latter gap is filled by
the Frobenius-Perron operator, which provides a rule to determine
how the evolution of densities over repeated iterations is
accomplished.

In the present paper we adopt the Renormalization Group (RG)
method for a constructive analysis of the properties of the
Frobenius-Perron Operator. The basic idea of the method is to
absorb secular or divergent terms of the naive perturbation
solution into renormalized integration constants (amplitudes). As
a result one obtains an evolution law embedded in an evolution
equation the renormalized amplitudes must satisfy, which describes
the large-scale dynamics of the system. The latter is usually
called the RG equation. The stages in the renormalization group
reduction of a particular physical system are quite general and
well defined, which makes the RG method universal and independent
on the concrete details of the underlying dynamics and physical
processes involved. \cite{Tzenov,Chen,Oono,Kunihiro,Goto}

The paper is organized as follows. The next paragraph serves as a
reminder of the basic definition, derivation and properties of the
Frobenius-Perron operator and provides the starting point for the
subsequent exposition. In Paragraph 3, we work out in detail the
renormalization group reduction of a generic symplectic map in the
case, where the unperturbed rotation frequency of the map is far
from structural resonances driven by the kick perturbation.
Paragraph 4 deals with the resonance structure of a symplectic
map. It is shown that in the case, where the unperturbed rotation
frequency is close to a resonance, the reduced RG map of the
Frobenius-Perron operator is equivalent to a discrete
Fokker-Planck equation for the renormalized distribution function.
In Paragraph 5 we apply the RG method to study the stochastic
properties of the standard Chirikov-Taylor map, and (re)derive the
diffusion coefficient in the quasi-linear approximation. In
Paragraph 6 a few examples concerning the application of the
results thus obtained are worked out in detail. Finally, Paragraph
7 is dedicated to conclusions and outlook.

\section{The Frobenius-Perron Operator for the Henon Map}
The Henon map is defined by the following expression
\cite{Tzenov}:
\begin{equation}
{\bf z}_{n+1} = {\left(
\begin{array}{c}
x_{n+1} \\
p_{n+1}
\end{array} \right)} = {\cal R}_{\omega} {\left(
\begin{array}{c}
x_n \\
p_n - {\cal S} x_n^2
\end{array} \right)}, \label{HenonMap}
\end{equation}
\noindent where
\begin{equation}
{\cal R}_{\omega} = {\left(
\begin{array}{c}
\cos \omega \ \sin \omega \\
- \sin \omega \  \cos \omega
\end{array} \right)}, \label{RotMatrix}
\end{equation}
\noindent is the rotation matrix for one period of the map. In
terms of accelerator physics application this is equivalent to one
revolution along the accelerator lattice. The frequency $\omega$
and the parameter ${\cal S}$
\begin{equation}
\omega = 2 \pi \nu, \qquad \qquad {\cal S} = {\frac {l \lambda_0
{\left( \theta_0 \right)} \beta^{3/2} {\left( \theta_0 \right)}}
{2 R^3}}, \label{TuneParam}
\end{equation}
\noindent are related to the unperturbed betatron tune $\nu$ and
to the strength of the sextupole (cubic nonlinearity) perturbation
$\lambda_0$. Here $l$ is the length of the sextupole, $\theta_0$
is its location on the azimuth of the machine and $R$ is the mean
radius.

The Henon map can be written as
\begin{equation}
{\bf Z}_{n+1} = {\cal R}_{\omega}^T {\bf z}_{n+1} = {\left(
\begin{array}{c}
x_n \\
p_n - {\cal S} x_n^2
\end{array} \right)}, \label{HenonMalt}
\end{equation}
\noindent where ${\cal R}_{\omega}^T$ denotes the transposed of
the matrix (\ref{RotMatrix}). The Frobenius-Perron operator
\cite{Tzenov} can be calculated explicitly. We have:
\begin{eqnarray}
f_{n+1} {\left( x, p \right)} = {\widehat{\bf U}} f_n {\left( x, p
\right)} = \int {\rm d} \xi {\rm d} \eta \delta {\left( X - \xi
\right)} \delta {\left( P - \eta + {\cal S} \xi^2 \right)} f_n
{\left( \xi, \eta \right)} \nonumber
\end{eqnarray}
\begin{equation}
= f_n {\left( X, P + {\cal S} X^2 \right)}. \label{FrobPerroper}
\end{equation}
\noindent Introducing the formal small parameter $\epsilon$ and
the action-angle variables
\begin{equation}
x = {\sqrt{2J}} \cos a, \qquad \qquad p = - {\sqrt{2J}} \sin a,
\label{ActionAngle}
\end{equation}
\noindent with
\begin{equation}
J = {\frac {1} {2}} {\left( x^2 + p^2 \right)}, \qquad \qquad a =
- \arctan {\left( {\frac {p} {x}} \right)}, \label{ActionAngle1}
\end{equation}
\noindent we write the Frobenius-Perron operator represented by
equation (\ref{FrobPerroper}) in the form
\begin{equation}
f_{n+1} {\left( a + \omega, J \right)} = f_n {\left( x, p +
\epsilon {\cal S} x^2 \right)}. \label{FrobPerron}
\end{equation}
\noindent

\section{Renormalization Group Treatment of the Frobenius-Perron
Operator}

The generalization of the Frobenius-Perron operator
(\ref{FrobPerron}) for a generic symplectic map with rotation is
straightforward. We have
\begin{equation}
f_{n+1} {\left( a + \omega, J \right)} = f_n {\left( x, p +
\epsilon \partial_x V_N \right)}, \label{FrobPerrons}
\end{equation}
\noindent where $V_N {\left( x \right)}$ is a potential and
$\partial_x$ denotes partial differentiation with respect to $x$.
In the case of the Henon map for instance, the potential has the
form
\begin{equation}
V_N {\left( x \right)} = {\frac {{\cal S} x^3} {3}}.
\label{PotentHenon}
\end{equation}
\noindent Equation (\ref{FrobPerrons}) can be written as
\begin{equation}
f_{n+1} {\left( a + \omega, J \right)} = e^{\epsilon {\left(
\partial_x V_N \right)} \partial_p} f_n {\left( a, J \right)}.
\label{FrobPerronex}
\end{equation}
\noindent Since the potential $V_N$ does not depend on the
momentum variable $p$,
\begin{equation}
{\widehat{\bf L}}_V = {\left( \partial_x V_N \right)} \partial_p -
{\left( \partial_p V_N \right)} \partial_x = {\left( \partial_x
V_N \right)} \partial_p, \label{Liouvillean}
\end{equation}
\noindent where ${\widehat{\bf L}}_V$ is the Liouvillean operator
associated with $V_N$. Therefore, equation (\ref{FrobPerronex})
becomes
\begin{equation}
f_{n+1} {\left( a + \omega, J \right)} = e^{\epsilon {\widehat{\bf
L}}_V} f_n {\left( a, J \right)}. \label{FrobPerroex}
\end{equation}
\noindent We assume that the potential $V_N$, written in
action-angle variables can be split as follows
\begin{equation}
V_N {\left( a, J \right)} = V_0 {\left( J \right)} + V {\left( a,
J \right)}. \label{Potential}
\end{equation}
\noindent Respectively, the Liouvillean operator can be written as
\begin{equation}
{\widehat{\bf L}}_V = {\widehat{\bf L}}_0 + {\widehat{\bf L}},
\label{Liouvillop}
\end{equation}
\noindent where
\begin{equation}
{\widehat{\bf L}}_0 = - \omega_V {\left( J \right)} \partial_a,
\qquad \qquad {\widehat{\bf L}} = {\left( \partial_a V \right)}
\partial_J - {\left( \partial_J V \right)} \partial_a, \label{Liouvillope}
\end{equation}
\noindent and
\begin{equation}
\omega_V {\left( J \right)} = {\frac {\partial V_0} {\partial J}}.
\label{Tuneshift}
\end{equation}

First of all, we consider the case, where the rotation frequency
$\omega$ is away from nonlinear resonances driven by the potential
$V$. Following the standard procedure of the RG method
\cite{Tzenov,Chen}, we seek a solution to equation
(\ref{FrobPerroex}) by naive perturbation expansion
\begin{equation}
f_n {\left( a, J \right)} = \sum \limits_{k=0}^{\infty} \epsilon^k
f_n^{(k)} {\left( a, J \right)}, \label{Naivepertur}
\end{equation}
\noindent where the unknown functions $f_n^{(k)} {\left( a, J
\right)}$ have to be determined order by order. The zero-order
equation
\begin{equation}
f_{n+1}^{(0)} {\left( a + \omega, J \right)} = f_n^{(0)} {\left(
a, J \right)}, \label{Zeroordequ}
\end{equation}
\noindent has the obvious solution
\begin{equation}
f_n^{(0)} {\left( a, J \right)} = e^{ - n \omega \partial_a} F
{\left( a, J \right)} = F {\left( a - n \omega, J \right)}.
\label{Zeroordsolu}
\end{equation}
\noindent To this end $F {\left( a, J \right)}$ is an arbitrary
function of its arguments, and will be the subject of the
renormalization group reduction in the sequel.

The first-order equation can be written as follows
\begin{equation}
f_{n+1}^{(1)} {\left( a + \omega, J \right)} - f_n^{(1)} {\left(
a, J \right)} = {\widehat{\bf L}}_V F {\left( a - n \omega, J
\right)}. \label{Firstordequ}
\end{equation}
\noindent Standard but cumbersome algebra yields the solution to
equation (\ref{Firstordequ}) in the form
\begin{equation}
f_n^{(1)} {\left( a, J \right)} = {\left( n {\widehat{\bf L}}_0 +
{\widehat{\cal L}}_{\omega} \right)} F {\left( a - n \omega, J
\right)}, \label{Firstordsolu}
\end{equation}
\noindent where
\begin{equation}
{\widehat{\cal L}}_{\omega} = {\left( \partial_a V_{\omega}
\right)} \partial_J - {\left( \partial_J V_{\omega} \right)}
\partial_a. \label{Liouvilleom}
\end{equation}
\noindent Furthermore, the potential $V_{\omega} {\left( a, J
\right)}$ is defined according to the expression
\begin{equation}
V_{\omega} {\left( a, J \right)} = V_1 {\left( a - {\frac {\omega}
{2}}, J \right)}, \qquad \qquad V_1 {\left( a, J \right)} = \sum
\limits_{m \neq 0} {\frac {V_m {\left( J \right)} e^{ima}} {2i
\sin {\left( m \omega / 2 \right)}}}. \label{Potentialom}
\end{equation}
\noindent Some of the details of the calculation can be found in
Appendix A.

The second order equation is
\begin{equation}
f_{n+1}^{(2)} {\left( a + \omega, J \right)} - f_n^{(2)} {\left(
a, J \right)} = {\widehat{\bf L}}_V f_n^{(1)} {\left( a, J
\right)} + {\frac {{\widehat{\bf L}}_V^2} {2}} F {\left( a - n
\omega, J \right)}. \label{Secondordequ}
\end{equation}
\noindent Since we are interested in the secular solution of
equation (\ref{Secondordequ}), we retain on its right-hand-side
only terms that would yield a secular contribution. Thus, the
second order equation giving rise to secular solution can be
written as
\begin{equation}
f_{n+1}^{(2)} {\left( a + \omega, J \right)} - f_n^{(2)} {\left(
a, J \right)} = {\left[ {\left( n + {\frac {1} {2}} \right)}
{\widehat{\bf L}}_0^2 + n {\widehat{\bf L}} {\widehat{\bf L}}_0 +
\Omega {\left( \omega, J \right)} \partial_a \right]} F {\left( a
- n \omega, J \right)}, \label{Secondordequs}
\end{equation}
\noindent where
\begin{equation}
\Omega {\left( \omega, J \right)} = \sum \limits_{m=1}^{\infty} m
\cot {\left( {\frac {m \omega} {2}} \right)} \partial_J {\left(
V_m \partial_J V_m \right)}. \label{Omega}
\end{equation}
\noindent Omitting the details of the calculation (presented in
Appendix A), we can write the second-order solution as
\begin{equation}
f_n^{(2)} {\left( a, J \right)} = {\left[ {\frac {n^2} {2}}
{\widehat{\bf L}}_0^2 + n {\widehat{\cal L}}_{\omega}
{\widehat{\bf L}}_0 + n \Omega {\left( \omega, J \right)}
\partial_a \right]} F {\left( a - n \omega, J \right)} + {\rm non}
\; {\rm secular} \; {\rm terms}. \label{Secondordsols}
\end{equation}

To remove secular terms (proportional to $n$ and $n^2$) in the
first-order (\ref{Firstordsolu}) and the second-order solution
(\ref{Secondordsols}), we define a renormalization group
transformation $F {\left( a, J \right)} \rightarrow
{\widetilde{F}} {\left( a, J; n \right)}$ by collecting all terms
proportional to $F {\left( a - n \omega, J \right)}$
\begin{equation}
{\widetilde{F}} {\left( a - n \omega, J; n \right)} = {\left[ 1 +
\epsilon n {\widehat{\bf L}}_0 + \epsilon^2 {\left( {\frac {n^2}
{2}} {\widehat{\bf L}}_0^2 + n \Omega \partial_a \right)} \right]}
F {\left( a - n \omega, J \right)}. \label{Renormtransform}
\end{equation}
\noindent Solving perturbatively equation (\ref{Renormtransform})
for $F {\left( a - n \omega, J \right)}$ in terms of
${\widetilde{F}} {\left( a - n \omega, J; n \right)}$, we obtain
\begin{equation}
F {\left( a - n \omega, J \right)} = {\left( 1 - \epsilon n
{\widehat{\bf L}}_0 + \dots \right)} {\widetilde{F}} {\left( a - n
\omega, J; n \right)}. \label{Renormtransf}
\end{equation}
\noindent Following Reference \cite{Tzenov,Goto}, we define a
discrete version of the RG equation by considering the difference
\begin{equation}
{\widetilde{F}} {\left( a - n \omega, J; n+1 \right)} -
{\widetilde{F}} {\left( a - n \omega, J; n \right)} \nonumber
\end{equation}
\begin{equation}
= {\left\{ \epsilon {\widehat{\bf L}}_0 + \epsilon^2 {\left[
{\left( n + {\frac {1} {2}} \right)} {\widehat{\bf L}}_0^2 +
\Omega \partial_a \right]} \right\}} F {\left( a - n \omega, J
\right)}. \label{Renormequat}
\end{equation}
\noindent Substituting the expression for $F {\left( a - n \omega,
J \right)}$ in terms of ${\widetilde{F}} {\left( a - n \omega, J;
n \right)}$ from equation (\ref{Renormtransf}), we can eliminate
secular terms up to $O {\left( \epsilon^2 \right)}$. The result is
\begin{equation}
{\widetilde{F}} {\left( a - n \omega, J; n+1 \right)} -
{\widetilde{F}} {\left( a - n \omega, J; n \right)} = {\left[
\epsilon {\widehat{\bf L}}_0 + \epsilon^2 {\left( {\frac
{{\widehat{\bf L}}_0^2} {2}} + \Omega \partial_a \right)} \right]}
{\widetilde{F}} {\left( a - n \omega, J; n \right)}.
\label{Renormequation}
\end{equation}
\noindent Equation (\ref{Renormequation}) is the RG equation. It
describes the evolution of the distribution function on slow time
scales in addition to the fast oscillations with a fundamental
frequency $\omega$.

An important remark is in order at this point. Note that once the
renormalization transformation has been performed, the second term
in the second-order solution (\ref{Secondordsols}) is eliminated
as well. Combining it with the second (non secular) term in the
first-order solution (\ref{Firstordsolu}), we obtain
\begin{equation}
\epsilon {\widehat{\cal L}}_{\omega} F + \epsilon^2 n
{\widehat{\cal L}}_{\omega} {\widehat{\bf L}}_0 F = \epsilon
{\widehat{\cal L}}_{\omega} {\left( 1 - \epsilon n {\widehat{\bf
L}}_0 \right)} {\widetilde{F}} {\left( n \right)} + \epsilon^2 n
{\widehat{\cal L}}_{\omega} {\widehat{\bf L}}_0 {\widetilde{F}}
{\left( n \right)} = \epsilon {\widehat{\cal L}}_{\omega}
{\widetilde{F}} {\left( n \right)}. \label{Cancellat}
\end{equation}

To first order in the perturbation parameter $\epsilon$ the
renormalized solution to equation (\ref{FrobPerroex}) can be
written as
\begin{equation}
f_n {\left( a, J \right)} = {\left( 1 + \epsilon {\widehat{\cal
L}}_{\omega} \right)} {\widetilde{F}} {\left( a - n \omega, J; n
\right)}, \label{Renormsol}
\end{equation}
\noindent where the renormalized "amplitude" ${\widetilde{F}}
{\left( a - n \omega, J; n \right)}$ satisfies the RG equation
(\ref{Renormequation}). In the continuous limit equation
(\ref{Renormequation}) acquires the form
\begin{equation}
{\frac {\partial {\widetilde{F}} {\left( a - n \omega, J; n
\right)}} {\partial n}} = {\left[ \epsilon {\widehat{\bf L}}_0 +
\epsilon^2 {\left( {\frac {{\widehat{\bf L}}_0^2} {2}} + \Omega
\partial_a \right)} \right]} {\widetilde{F}} {\left( a - n \omega,
J; n \right)}. \label{Renormequationc}
\end{equation}
\noindent Provided ${\widehat{\bf L}}_0 \neq 0$ (in the case,
where the potential $V_N$ is not antisymmetric) the latter is a
Fokker-Planck equation with the Fokker-Planck operator acting on
the angle variable only. A relevant example of a cubic map will be
considered in Paragraph 6.

\section{Resonance Structure of a Symplectic Map}

The solution (\ref{Firstordsolu}) to the first-order perturbation
equation (\ref{Firstordequ}) was obtained under the assumption
that the unperturbed betatron tune $\nu$ is sufficiently far from
any structural nonlinear resonance of the form $m_0 \nu = 1$,
where $m_0$ is an integer. In the present paragraph, we assume
that
\begin{equation}
\omega = \omega_0 + \epsilon \delta_1 + \epsilon^2 \delta_2 +
\dots, \qquad \qquad \omega_0 = {\frac {2 \pi} {m_0}}.
\label{RotationFreq}
\end{equation}
\noindent Moreover, for the sake of simplicity, we assume that
there are no higher angle-dependent harmonics in the Fourier
spectrum of $V {\left( a, J \right)}$ that would drive
higher-order resonances of the form $p m_0 \nu = p$, where $p$ is
an integer. However, results can be generalized easily to take
into account this case as well.

Proceeding as in Paragraph 3, we write the zero-order solution as
\begin{equation}
f_n^{(0)} {\left( a, J \right)} = e^{ - n \omega_0 \partial_a} F
{\left( a, J \right)} = F {\left( a - n \omega_0, J \right)}.
\label{Zeroordsolur}
\end{equation}
\noindent The first-order equation in the resonant case can be
written as follows
\begin{equation}
f_{n+1}^{(1)} {\left( a + \omega_0, J \right)} - f_n^{(1)} {\left(
a, J \right)} = {\left( {\widehat{\bf L}}_1 + {\widehat{\bf L}}
\right)} F {\left( a - n \omega_0, J \right)},
\label{Firstordequr}
\end{equation}
\noindent where
\begin{equation}
{\widehat{\bf L}}_1 = - \delta_1 \partial_a + {\widehat{\bf L}}_0.
\label{OperatorL1}
\end{equation}
\noindent The solution to equation (\ref{Firstordequr}) is readily
found in a standard manner, analogous to that already used in
Paragraph 3 (see also Appendix A). The result is
\begin{equation}
f_n^{(1)} {\left( a, J \right)} = {\left[ n {\left( {\widehat{\bf
L}}_1 + {\widehat{\bf L}}_R \right)} + {\widehat{\cal
L}}_{\omega}^{\prime} \right]} F {\left( a - n \omega_0, J
\right)}, \label{Firstordsolur}
\end{equation}
\noindent where
\begin{equation}
{\widehat{\bf L}}_R = {\left( \partial_a V_R \right)} \partial_J -
{\left( \partial_J V_R \right)} \partial_a, \label{Liouvilleres}
\end{equation}
\noindent is the resonant Liouvillean operator and
\begin{equation}
V_R {\left( a, J \right)} = \sum \limits_{m = \pm m_0} V_m {\left(
J \right)} e^{ima} = 2 V_{m0} {\left( J \right)} \cos m_0 a,
\label{Potentialres}
\end{equation}
\noindent is the resonant potential. Furthermore,
\begin{equation}
{\widehat{\cal L}}_{\omega}^{\prime} = {\left( \partial_a
V_{\omega}^{\prime} \right)} \partial_J - {\left( \partial_J
V_{\omega}^{\prime} \right)} \partial_a, \label{Liouvilleomr}
\end{equation}
\noindent where now the potential $V_{\omega}^{\prime} {\left( a,
J \right)}$ is defined according to the expression
\begin{equation}
V_{\omega}^{\prime} {\left( a, J \right)} = V_1^{\prime} {\left( a
- {\frac {\omega_0} {2}}, J \right)}, \qquad \qquad V_1^{\prime}
{\left( a, J \right)} = \sum \limits_{m \neq \pm m_0} {\frac {V_m
{\left( J \right)} e^{ima}} {2i \sin {\left( m \omega_0 / 2
\right)}}}. \label{Potentialomr}
\end{equation}

The second-order equation in the resonant case can be written as
\begin{eqnarray}
f_{n+1}^{(2)} {\left( a + \omega_0, J \right)} - f_n^{(2)} {\left(
a, J \right)} = - \delta_1 \partial_a f_{n+1}^{(1)} {\left( a +
\omega_0, J \right)} \nonumber
\end{eqnarray}
\begin{equation}
- {\frac {1} {2}} {\left( \delta_1^2
\partial^2_{aa} + 2 \delta_2 \partial_a \right)} f_{n+1}^{(0)}
{\left( a + \omega_0, J \right)} + {\widehat{\bf L}}_V f_n^{(1)}
{\left( a, J \right)} + {\frac {{\widehat{\bf L}}_V^2} {2}} F
{\left( a - n \omega_0, J \right)}. \label{Secondordequr}
\end{equation}
\noindent Similar to Paragraph 3, we again retain terms on the
right-hand-side of equation (\ref{Secondordequr}) that would yield
secular contributions to the second-order solution. Thus the
second-order solution can be written as
\begin{eqnarray}
f_n^{(2)} {\left( a, J \right)} = n {\left\{ {\left(
\Omega^{\prime} - \delta_2 \right)} \partial_a + {\frac {\delta_1}
{2}} {\left[ {\widehat{\bf L}}_R, \partial_a \right]} \right\}} F
{\left( a - n \omega_0, J \right)} \nonumber
\end{eqnarray}
\begin{equation}
+ {\frac {n^2} {2}} {\left( {\widehat{\bf L}}_1 + {\widehat{\bf
L}}_R  \right)}^2 F {\left( a - n \omega_0, J \right)} + n
{\widehat{\cal L}}_{\omega}^{\prime} {\left( {\widehat{\bf L}}_1 +
{\widehat{\bf L}}_R \right)} F {\left( a - n \omega_0, J \right)}
+ {\rm non} \; {\rm secular} \; {\rm terms},
\label{Secondordsolsr}
\end{equation}
\noindent where
\begin{equation}
{\left[ {\widehat{\bf L}}_R, \partial_a \right]} = {\widehat{\bf
L}}_R \partial_a - \partial_a {\widehat{\bf L}}_R,
\label{Commutator}
\end{equation}
\noindent is the commutator of the operators ${\widehat{\bf L}}_R$
and $\partial_a$, and
\begin{equation}
\Omega^{\prime} {\left( \omega_0, J \right)} = {\sum
\limits_{m=1}^{\infty}}^{\prime} m \cot {\left( {\frac {m
\omega_0} {2}} \right)} \partial_J {\left( V_m \partial_J V_m
\right)}. \label{Omegap}
\end{equation}
\noindent is the new nonlinear tune shift. Here the prime in the
above sum implies that the term with $m = m_0$ is excluded from
the sum.

Repeating the steps that brought us along from equation
(\ref{Renormtransform}) to equation (\ref{Renormequation}) in the
previous paragraph, we obtain the RG equation in the resonant case
\begin{eqnarray}
{\widetilde{F}} {\left( a_n, J; n+1 \right)} - {\widetilde{F}}
{\left( a_n, J; n \right)} \nonumber
\end{eqnarray}
\begin{equation}
= {\left\{ \epsilon {\left( {\widehat{\bf L}}_1 + {\widehat{\bf
L}}_R \right)} + \epsilon^2 {\left[ {\left( \Omega^{\prime} -
\delta_2 \right)} \partial_a + {\frac {\delta_1} {2}} {\left[
{\widehat{\bf L}}_R, \partial_a \right]} + {\frac {1} {2}} {\left(
{\widehat{\bf L}}_1 + {\widehat{\bf L}}_R \right)}^2 \right]}
\right\}} {\widetilde{F}} {\left( a_n, J; n \right)},
\label{Renormequationr}
\end{equation}
\noindent where $a_n = a - n \omega_0$. Analogously to Paragraph
3, to first order in the perturbation parameter $\epsilon$ the
renormalized solution of equation (\ref{FrobPerroex}) in the
resonant case is represented by the expression
\begin{equation}
f_n {\left( a, J \right)} = {\left( 1 + \epsilon {\widehat{\cal
L}}_{\omega}^{\prime} \right)} {\widetilde{F}} {\left( a - n
\omega_0, J; n \right)}, \label{Renormsolres}
\end{equation}

As in the non resonant case, the renormalization group
transformation ensures the elimination of the last term on the
right-hand-side of equation (\ref{Secondordsolsr}) [combined with
the last term on the right-hand-side of equation
(\ref{Firstordsolur})] up to $O {\left( \epsilon^2 \right)}$. In
the continuous limit the RG equation (\ref{Renormequationr}) is a
Fokker-Planck equation with a Fokker-Planck operator acting on
both action and angle variables.

\section{Stochastic Properties of the Standard Map}

The present paragraph is dedicated to the analysis of the
stochastic properties of the standard map. The Frobenius-Perron
operator can be calculated explicitly and the Frobenius-Perron
equation can be written as \cite{Tzenov}
\begin{equation}
f_{n+1} {\left( a, J \right)} = e^{- J \partial_a} e^{\epsilon
\sin a \partial_J} f_n {\left( a, J \right)}, \label{FPequatsm1}
\end{equation}
\noindent or
\begin{equation}
e^{J \partial_a} f_{n+1} {\left( a, J \right)} = e^{\epsilon \sin
a \partial_J} f_n {\left( a, J \right)}. \label{FPequatsm2}
\end{equation}
\noindent We introduce a new variable $\xi$ according to the
equation
\begin{equation}
\xi = {\frac {J - J_0} {\sqrt{\epsilon}}}, \qquad \qquad J = J_0 +
{\sqrt{\epsilon}} \xi, \label{Xivariable}
\end{equation}
\noindent where $J_0$ is a fixed value of the action variable. The
first case we will consider is the one, where $J_0 \neq 0$. The
Frobenius-Perron equation (\ref{FPequatsm2}) can be rewritten as
\begin{equation}
e^{{\left( J_0 + {\sqrt{\epsilon}} \xi \right)} \partial_a}
f_{n+1} {\left( a, \xi \right)} = e^{{\sqrt{\epsilon}} \sin a
\partial_{\xi}} f_n {\left( a, \xi \right)}. \label{FPequatsm}
\end{equation}
\noindent The zero-order equation
\begin{equation}
e^{J_0 \partial_a} f_{n+1}^{(0)} {\left( a, \xi \right)} =
f_{n+1}^{(0)} {\left( a + J_0, \xi \right)} = f_n^{(0)} {\left( a,
\xi \right)}, \label{Zeroordequsm}
\end{equation}
\noindent has the obvious solution
\begin{equation}
f_n^{(0)} {\left( a, \xi \right)} = e^{ - n J_0 \partial_a} F
{\left( a, \xi \right)} = F {\left( a - n J_0, \xi \right)}.
\label{Zeroordsolusm}
\end{equation}

The first-order perturbation equation can be written as follows
\begin{equation}
e^{J_0 \partial_a} f_{n+1}^{(1)} {\left( a, \xi \right)} -
f_n^{(1)} {\left( a, \xi \right)} = {\left( -\xi \partial_a + \sin
a \partial_{\xi} \right)} F {\left( a - n J_0, \xi \right)}.
\label{Firstordequsm}
\end{equation}
\noindent The latter can be solved in a straightforward manner and
the result is
\begin{equation}
f_n^{(1)} {\left( a, \xi \right)} = - {\left[ n \xi \partial_a +
{\frac {\cos {\left( a - J_0 / 2 \right)}} {2 \sin {\left( J_0 / 2
\right)}}} \partial_{\xi} \right]} F {\left( a - n J_0, \xi
\right)}. \label{Firstordsolsm}
\end{equation}
\noindent Proceeding further, we write the second-order
perturbation equation
\begin{equation}
e^{J_0 \partial_a} f_{n+1}^{(2)} - f_n^{(2)} = - e^{J_0
\partial_a} {\left( \xi \partial_a f_{n+1}^{(1)} + {\frac {\xi^2}
{2}} \partial_{aa}^2 f_{n+1}^{(0)} \right)} + \sin a
\partial_{\xi} f_n^{(1)} + {\frac {1} {2}} \sin^2 a \partial_{\xi
\xi}^2 f_n^{(0)}. \label{Secondordequsm}
\end{equation}
\noindent Retaining terms on the right-hand-side of equation
(\ref{Secondordequsm}) that would give rise to secular
contribution, we cast the latter in an explicit form according to
the expression
\begin{equation}
e^{J_0 \partial_a} f_{n+1}^{(2)} - f_n^{(2)} = \xi^2 {\left( n +
{\frac {1} {2}} \right)} \partial_{aa}^2 F {\left( a - n J_0, \xi
\right)} - n \sin a \partial_{\xi} \xi \partial_a F {\left( a - n
J_0, \xi \right)} \nonumber
\end{equation}
\begin{equation}
- {\frac {\cot {\left( J_0 / 2 \right)}} {4}} \sin 2a
\partial_{\xi \xi}^2 F {\left( a - n J_0, \xi \right)}.
\label{Secondordequsme}
\end{equation}

The secular solution of equation (\ref{Secondordequsme}) is
\begin{equation}
f_n^{(2)} {\left( a, \xi \right)} = {\frac {n^2} {2}} \xi^2
\partial_{aa}^2 F {\left( a - n J_0, \xi \right)} + n {\frac {\cos
{\left( a - J_0 / 2 \right)}} {2 \sin {\left( J_0 / 2 \right)}}}
\partial_{\xi} \xi \partial_a F {\left( a - n J_0, \xi \right)}.
\label{Secondordsolsm}
\end{equation}
\noindent Following out a renormalization procedure similar to the
one performed in Paragraphs 3 and 4, we obtain the RG equation
\begin{equation}
{\widetilde{F}} {\left( a - n J_0, \xi; n+1 \right)} -
{\widetilde{F}} {\left( a - n J_0, \xi; n \right)} = {\left( -
{\sqrt{\epsilon}} \xi \partial_a + {\frac {\epsilon \xi^2} {2}}
\partial_{aa}^2 \right)} {\widetilde{F}} {\left( a - n J_0, \xi; n
\right)}. \label{Renormequatsmnonr}
\end{equation}
\noindent Note that in the case of $J_0 = \pi$, the third term on
the right-hand-side of equation (\ref{Secondordequsme}) vanishes
and we are again left with the above expressions for the secular
second-order solution (\ref{Secondordsolsm}) and for the
renormalization group map (\ref{Renormequatsmnonr}) (taken for
$J_0 = \pi$).

To complete the present paragraph, we consider the case of $J_0 =
0$. Equation (\ref{Xivariable}) implies a simple (non canonical)
scaling of the canonical variables, assuming that $J$ is a slow
variable. Then, equation (\ref{FPequatsm1}) can be rewritten as
\begin{equation}
f_{n+1} {\left( a, \xi \right)} = e^{- {\sqrt{\epsilon}} \xi
\partial_a} e^{{\sqrt{\epsilon}} \sin a \partial_{\xi}} f_n {\left( a,
\xi \right)}. \label{FPequatsmsc1}
\end{equation}
\noindent Note that now the perturbation parameter is
${\sqrt{\epsilon}}$ rather than $\epsilon$. Using the
Campbell-Baker-Hausdorff identity
\begin{equation}
\exp {\left( \alpha {\widehat{\bf A}} \right)} \exp {\left( \beta
{\widehat{\bf B}} \right)} = \exp {\left( \alpha {\widehat{\bf A}}
+ \beta {\widehat{\bf B}} + {\frac {\alpha \beta} {2}} {\left[
{\widehat{\bf A}}, {\widehat{\bf B}} \right]} + \dots \right)},
\label{CBPident}
\end{equation}
\noindent for any operators ${\widehat{\bf A}}$ and ${\widehat{\bf
B}}$ and any parameters $\alpha$ and $\beta$, we cast equation
(\ref{FPequatsmsc1}) in the form
\begin{equation}
f_{n+1} {\left( a, \xi \right)} = \exp {\left( {\sqrt{\epsilon}}
{\widehat{\bf L}}_S + {\frac {\epsilon} {2}} {\widehat{\bf C}} +
\dots \right)} f_n {\left( a, \xi \right)}. \label{FPequatsmsc}
\end{equation}
\noindent Here
\begin{equation}
{\widehat{\bf L}}_S = - \xi \partial_a + \sin a \partial_{\xi},
\qquad \qquad {\widehat{\bf C}} = {\left[ \sin a \partial_{\xi},
\xi \partial_a \right]} = \sin a \partial_a - \xi \cos a
\partial_{\xi}. \label{LSoperat}
\end{equation}
\noindent Next we proceed with solving of equation
(\ref{FPequatsmsc}) order by order.

The zero-order solution is trivial to find:
\begin{equation}
f_n^{(0)} {\left( a, \xi \right)} = F {\left( a, \xi \right)},
\label{FPzerosolsmsc}
\end{equation}
\noindent where as before, $F {\left( a, \xi \right)}$ is a
generic function of its arguments, playing the role of an
integration constant. The first-order equation
\begin{equation}
f_{n+1}^{(1)} {\left( a, \xi \right)} - f_n^{(1)} {\left( a, \xi
\right)} = {\widehat{\bf L}}_S F {\left( a, \xi \right)},
\label{FPfoequatsmsc}
\end{equation}
\noindent has the obvious solution
\begin{equation}
f_n^{(1)} {\left( a, \xi \right)} = n {\widehat{\bf L}}_S F
{\left( a, \xi \right)}. \label{FPfosolsmsc}
\end{equation}
\noindent The second-order equation can be written explicitly as
\begin{equation}
f_{n+1}^{(2)} {\left( a, \xi \right)} - f_n^{(2)} {\left( a, \xi
\right)} = {\left[ {\left( n + {\frac {1} {2}} \right)}
{\widehat{\bf L}}_S^2 + {\frac {{\widehat{\bf C}}} {2}} \right]} F
{\left( a, \xi \right)}. \label{FPsoequatsmsc}
\end{equation}
\noindent Its solution is readily found to be
\begin{equation}
f_n^{(2)} {\left( a, \xi \right)} = {\frac {1} {2}} {\left( n^2
{\widehat{\bf L}}_S^2 + n {\widehat{\bf C}} \right)} F {\left( a,
\xi \right)}. \label{FPsosolsmsc}
\end{equation}
\noindent Performing the already familiar renormalization
transformation in analogy to what has been done in the previous
paragraphs, we obtain the RG equation
\begin{equation}
{\widetilde{F}} {\left( a, \xi; n+1 \right)} - {\widetilde{F}}
{\left( a, \xi; n \right)} = {\left[ {\sqrt{\epsilon}}
{\widehat{\bf L}}_S + {\frac {\epsilon} {2}} {\left( {\widehat{\bf
L}}_S^2 + {\widehat{\bf C}} \right)} \right]} {\widetilde{F}}
{\left( a, \xi; n \right)}. \label{Renormequatsmsc}
\end{equation}

Averaging the RG equation (\ref{Renormequatsmsc}) over the angle
variable $a$, we obtain a discrete version of the diffusion
equation
\begin{equation}
{\widetilde{F}}_0 {\left( \xi; n+1 \right)} - {\widetilde{F}}_0
{\left( \xi; n \right)} = {\frac {\epsilon} {4}} \partial_{\xi
\xi}^2 {\widetilde{F}}_0 {\left( \xi; n \right)} = {\frac
{\epsilon^2} {4}} \partial_{JJ}^2 {\widetilde{F}}_0 {\left( \xi; n
\right)}, \label{Renormequatsmql}
\end{equation}
\noindent for the angle-independent part of the distribution
function ${\widetilde{F}}_0 {\left( \xi; n \right)}$ in
quasi-linear approximation. Higher order contributions to the
diffusion coefficient $\epsilon^2 / 2$ can be obtained by direct
analysis of the Frobenius-Perron operator \cite{Tzenov}.

\section{Examples}

Let us now consider a few examples. In the non resonant case of
the Henon map ${\left( {\widehat{\bf L}}_0 \equiv 0 \right)}$
equation (\ref{Renormequationc}) can be written in the form
\begin{equation}
{\frac {\partial {\widetilde{F}} {\left( a - n \omega, J; n
\right)}} {\partial n}} = \Omega_H {\left( a, J \right)}
\partial_a {\widetilde{F}} {\left( a - n \omega, J; n \right)},
\label{Nonreshenon}
\end{equation}
\noindent where
\begin{equation}
\Omega_H {\left( \omega, J \right)} = {\frac {{\cal S}^2 J} {8}}
{\left[ 3 \cot {\left( {\frac {\omega} {2}} \right)} + \cot
{\left( {\frac {3 \omega} {2}} \right)} \right]}.
\label{Tuneshihenon}
\end{equation}
\noindent Equation (\ref{Nonreshenon}) written in alternative form
\begin{equation}
{\frac {\partial {\widetilde{F}} {\left( a, J; n \right)}}
{\partial n}} = {\left[ - \omega + \Omega_H {\left( a, J \right)}
\right]} \partial_a {\widetilde{F}} {\left( a, J; n \right)},
\label{Nonreshenona}
\end{equation}
\noindent describes regular motion with a frequency $\omega -
\Omega_H$, and effective Hamiltonian
\begin{equation}
H_{eff} {\left( J \right)} = \omega J - {\frac {{\cal S}^2} {16}}
{\left[ 3 \cot {\left( {\frac {\omega} {2}} \right)} + \cot
{\left( {\frac {3 \omega} {2}} \right)} \right]} J^2.
\label{Hameffhenon}
\end{equation}

The next example we would like to consider is the case of a cubic
map with a potential of the form
\begin{equation}
V_N {\left( x \right)} = {\frac {{\cal C} x^4} {4}}.
\label{Quartpot}
\end{equation}
\noindent Taking into account equation (\ref{Potential}), we also
have
\begin{equation}
V_0 {\left( J \right)} = {\frac {3 {\cal C} J^2} {8}}, \qquad
\qquad V {\left( a, J \right)} = {\frac {{\cal C} J^2} {2}} \cos
2a + {\frac {{\cal C} J^2} {8}} \cos 4a. \label{Quartpotent}
\end{equation}
\noindent Furthermore, the nonlinear tune shift can be expressed
according to
\begin{equation}
\omega_V {\left( J \right)} = {\frac {3 {\cal C} J} {4}}, \qquad
\qquad \Omega_C {\left( \omega, J \right)} = {\frac {3 {\cal C}^2
J^2} {32}} {\left( 8 \cot \omega + \cot 2 \omega \right)}.
\label{Tunes}
\end{equation}
\noindent Thus, equation (\ref{Renormequationc}) can be written as
\begin{equation}
{\frac {\partial {\widetilde{F}} {\left( a, J; n \right)}}
{\partial n}} = - {\widetilde{\omega}} \partial_a {\widetilde{F}}
{\left( a, J; n \right)} + {\frac {\omega_V^2} {2}}
\partial_{aa}^2 {\widetilde{F}} {\left( a, J; n \right)},
\label{Nonrescubica}
\end{equation}
\noindent where
\begin{equation}
{\widetilde{\omega}} {\left( \omega, J \right)} = \omega +
\omega_V - {\frac {3 {\cal C}^2 J^2} {32}} {\left( 8 \cot \omega +
\cot 2 \omega \right)}. \label{Tunetilde}
\end{equation}
\noindent Equation (\ref{Nonrescubica}) can be readily solved,
yielding the result
\begin{equation}
{\widetilde{F}} {\left( a, J; n \right)} = \sum \limits_{k}
{\widetilde{F}}_k {\left( J; 0 \right)} e^{ik {\left( a - n
{\widetilde{\omega}} \right)}} e^{- k^2 \omega_V^2 n / 2}.
\label{Diffeqsolut}
\end{equation}
\noindent The latter indicates that the renormalized distribution
function ${\widetilde{F}} {\left( a, J; n \right)}$ rapidly
relaxes towards the invariant density ${\widetilde{F}}_0 {\left( J
\right)}$.

To complete the present paragraph, we study the resonance case for
the Henon map. We assume that the unperturbed betatron tune is
close to a third order resonance $3 \nu_0 = 1$. For the operator
${\widehat{\bf L}}_1$ and for the nonlinear tune shift
$\Omega^{\prime} {\left( \omega_0, J \right)}$, we obtain
\begin{equation}
{\widehat{\bf L}}_1 = - \delta_1 \partial_a, \qquad \qquad
\Omega^{\prime} {\left( \omega_0, J \right)} = {\frac {{\sqrt{3}}
{\cal S}^2 J} {8}}, \label{OperatorhL1}
\end{equation}
\noindent respectively. In the continuous limit equation
(\ref{Renormequationr}) can be written as
\begin{equation}
{\frac {\partial {\widetilde{F}}} {\partial n}} = - {\left( \omega
- \Omega^{\prime} \right)} \partial_a {\widetilde{F}} +
{\widehat{\bf L}}_R {\widetilde{F}} + {\frac {1} {2}} {\left(
{\widehat{\bf L}}_1^2 + 2 {\widehat{\bf L}}_1 {\widehat{\bf L}}_R
+ {\widehat{\bf L}}_R^2 \right)} {\widetilde{F}},
\label{DiffHenon}
\end{equation}
\noindent where
\begin{equation}
{\widehat{\bf L}}_R = - {\frac {{\cal S} {\sqrt{2J}}} {4}} {\left[
2J \sin {\left( 3a \right)} \partial_J + \cos {\left( 3a \right)}
\partial_a \right]}, \label{OperatorhLr}
\end{equation}
\noindent the frequency $\omega$ is represented by expression
(\ref{RotationFreq}) up to $O {\left( \epsilon^2 \right)}$, and
the renormalized distribution function ${\widetilde{F}} =
{\widetilde{F}} {\left( a, J; n \right)}$ is a function of the
phase-space variables and the "time" $n$.

Equation (\ref{DiffHenon}) is a Fokker-Planck equation describing
the slow evolution of the phase-space density in the case where
the rotation frequency of the Henon map is close to a third order
resonance.

\section{Concluding Remarks}
We have applied the Renormalization Group (RG) method to study the
stochastic properties of the Frobenius-Perron operator for a
variety of symplectic maps. After a brief introduction and
derivation of the Frobenius-Perron operator for a generic
symplectic map with rotation, the case, where the unperturbed
rotation frequency of the map is far from structural resonances
driven by the kick perturbation has been analyzed in detail. It
has been shown that up to second order in the strength of the
perturbation kick, the renormalized propagator for maps with
nonlinear stabilization ${\left( {\widehat{\bf L}}_0 \neq 0
\right)}$ describes random wandering of the angle variable.
Further, the resonance structure of a symplectic map has been
investigated. It has been shown that in the case, where the
unperturbed rotation frequency is close to a resonance, the
reduced RG map of the Frobenius-Perron operator (or reduced
phase-space density propagator) is equivalent to a discrete
Fokker-Planck equation for the renormalized distribution function.

The RG method has been also applied to study the stochastic
properties of the standard Chirikov-Taylor map. A nontrivial
discrete analogue of the Fokker-Planck equation with a
Fokker-Planck operator acting on both canonical variables has been
obtained. The latter reduces to the well-known diffusion equation
in quasi-linear approximation for the angle-independent part of
the distribution function.

It is worthwhile to mention that the procedure developed in the
present paper (see Paragraph 4) can be applied with a slight
generalization to study modulational effects in symplectic maps.


\appendix
\section{} 
Writing equation (\ref{Firstordequ}) in the form
\begin{equation}
f_{n+1}^{(1)} {\left( a + \omega, J \right)} - f_n^{(1)} {\left(
a, J \right)} = {\left( {\widehat{\bf L}}_0 + {\widehat{\bf L}}
\right)} F {\left( a - n \omega, J \right)}, \label{Firstordequd}
\end{equation}
\noindent we notice that the right-hand-side will give rise to two
kinds of terms. Let us first consider the equation
\begin{equation}
\varphi_{n+1} {\left( a + \omega, J \right)} - \varphi_n {\left(
a, J \right)} = {\widehat{\bf L}}_0 F {\left( a - n \omega, J
\right)}. \label{Firstordequa}
\end{equation}
\noindent Taking into account the commutativity between
${\widehat{\bf L}}_0$ and $\partial_a$, we can rewrite equation
(\ref{Firstordequa}) as
\begin{equation}
e^{\omega \partial_a} \varphi_{n+1} {\left( a, J \right)} -
\varphi_n {\left( a, J \right)} = e^{- n \omega \partial_a}
{\widehat{\bf L}}_0 F {\left( a, J \right)}. \label{Firstordequaa}
\end{equation}
\noindent It is straightforward to verify that the solution of the
last equation is
\begin{equation}
\varphi_n {\left( a, J \right)} = n e^{- n \omega \partial_a}
{\widehat{\bf L}}_0 F {\left( a, J \right)} = n {\widehat{\bf
L}}_0 F {\left( a - n \omega, J \right)}. \label{Firstordsolaa}
\end{equation}

Since the potential $V {\left( a, J \right)}$, the arbitrary
function $F {\left( a, J \right)}$ and the first-order
distribution function $f_n^{(1)} {\left( a, J \right)}$ are
periodic in the angle variable $a$, we can represent them as a
Fourier series in $a$
\begin{equation}
V {\left( a, J \right)} = \sum \limits_{m \neq 0} V_m {\left( J
\right)} e^{ima}, \qquad \qquad F {\left( a, J \right)} = \sum
\limits_{s} F_s {\left( J \right)} e^{isa}, \label{FourierV}
\end{equation}
\begin{equation}
f_n^{(1)} {\left( a, J \right)} = \sum \limits_{k} G_k^{(n)}
{\left( J \right)} e^{ika}. \label{Fourierdf}
\end{equation}
\noindent We substitute the above expansions into both sides of
the remainder equation
\begin{equation}
\psi_{n+1} {\left( a + \omega, J \right)} - \psi_n {\left( a, J
\right)} = {\widehat{\bf L}} F {\left( a - n \omega, J \right)},
\label{Firstordequb}
\end{equation}
\noindent and after equating similar harmonics, we obtain
\begin{equation}
G_k^{(n+1)} e^{ik \omega} - G_k^{(n)} = \sum \limits_{m} {\left[
im V_m F_{k-m}^{\prime} - i (k-m) V_m^{\prime} F_{k-m} \right]}
e^{-i (k-m) n \omega}, \label{Fourierfirst}
\end{equation}
\noindent where the primes indicate differentiation with respect
to the action variable $J$. It is straightforward to verify that
the solution of equation (\ref{Fourierfirst}) has the form
\begin{equation}
G_k^{(n)} = \sum \limits_{m} {\left[ im {\frac {V_m e^{- im \omega
/ 2}} {2i \sin {\left( m \omega / 2 \right)}}} F_{k-m}^{\prime} -
i (k-m) {\frac {V_m^{\prime} e^{- im \omega / 2}} {2i \sin {\left(
m \omega / 2 \right)}}} F_{k-m} \right]} e^{-i (k-m) n \omega}.
\label{Fourierfisol}
\end{equation}
\noindent Substituting back expression (\ref{Fourierfisol}) into
the expansion (\ref{Fourierdf}) for the function $\psi_n$ and
rearranging terms, we obtain
\begin{equation}
\psi_n {\left( a, J \right)} = \sum \limits_{m,s} e^{ima} {\left[
im {\frac {V_m e^{- im \omega / 2}} {2i \sin {\left( m \omega / 2
\right)}}} F_s^{\prime} - is {\frac {V_m^{\prime} e^{- im \omega /
2}} {2i \sin {\left( m \omega / 2 \right)}}} F_s \right]} e^{is (a
- n \omega)}. \label{Fourierfosol}
\end{equation}

Since equation (\ref{Firstordequd}) is linear, its general
solution can be represented as a sum of expressions
(\ref{Firstordsolaa}) and (\ref{Fourierfosol}). To complete the
derivation of the first-order solution (\ref{Firstordsolu}), we
note that expression (\ref{Fourierfosol}) represents the Fourier
expansion of the second term in (\ref{Firstordsolu}), provided the
potential $V_{\omega} {\left( a, J \right)}$ is given by
expression (\ref{Potentialom}).

Here, we briefly sketch the derivation of equation
(\ref{Secondordsols}). The first and the last terms on the
right-hand-side of equation (\ref{Secondordequs}) can be treated
in a way analogous to the treatment of equation
(\ref{Firstordequa}). Consider the solution of the equation
\begin{equation}
\Psi_{n+1} {\left( a + \omega, J \right)} - \Psi_n {\left( a, J
\right)} = n {\widehat{\bf L}} {\widehat{\bf L}}_0 F {\left( a - n
\omega, J \right)}. \label{Secondordequsa}
\end{equation}
\noindent Using the representation (\ref{FourierV}) and
(\ref{Fourierdf}), we can write the solution of equation
(\ref{Secondordequsa}) as
\begin{equation}
\Psi_n {\left( a, J \right)} = \sum \limits_{k} {\cal G}_k^{(n)}
{\left( J \right)} e^{ika}. \label{Fourierdfpsi}
\end{equation}
\noindent It can be verified by direct substitution that the
functions ${\cal G}_k^{(n)}$ are given by the expression
\begin{equation}
{\cal G}_k^{(n)} = \sum \limits_{m} {\left( n A_{km} + B_{km}
\right)} e^{-i (k-m) n \omega} , \label{Fourierfisolpsi}
\end{equation}
\noindent where
\begin{equation}
A_{km} = im {\frac {V_m e^{- im \omega / 2}} {2i \sin {\left( m
\omega / 2 \right)}}} W_{k-m}^{\prime} - i (k-m) {\frac
{V_m^{\prime} e^{- im \omega / 2}} {2i \sin {\left( m \omega / 2
\right)}}} W_{k-m}, \label{CoefficA}
\end{equation}
\begin{equation}
B_{km} = im {\frac {V_m} {4 \sin^2 {\left( m \omega / 2 \right)}}}
W_{k-m}^{\prime} - i (k-m) {\frac {V_m^{\prime}} {4 \sin^2 {\left(
m \omega / 2 \right)}}} W_{k-m}. \label{CoefficB}
\end{equation}
\noindent Here
\begin{equation}
W {\left( a, J \right)} = {\widehat{\bf L}}_0 W {\left( a, J
\right)}. \label{FunctionW}
\end{equation}
\noindent Thus, the desired expression for the second term on the
right-hand-side of equation (\ref{Secondordsols}) is readily
obtained by taking into account the Fourier expansion of the
solution of equation (\ref{Secondordequsa}).

%

\end{document}